\documentclass[11pt, a4paper]{article}

\usepackage[utf8]{inputenc}
\usepackage[T1]{fontenc}
\usepackage[english]{babel}
\usepackage{lmodern} % Uses scalable Latin Modern fonts
\usepackage[a4paper, top=2.5cm, bottom=2.5cm, left=2cm, right=2cm]{geometry}
\usepackage{amsmath}      % For math equations
\usepackage{booktabs}     % For professional tables
\usepackage{graphicx}     % For handling images
\usepackage{float}        % For figure positioning
\usepackage{hyperref}     % For hyperlinks in references
\usepackage{caption}      % For better caption control
\usepackage{authblk}      % For author affiliations
\usepackage{enumitem}     % For list formatting

\hypersetup{
    colorlinks=true,
    linkcolor=blue,
    filecolor=magenta,      
    urlcolor=cyan,
    citecolor=blue,
}

\title{\textbf{Evaluation Framework for Centralized and Decentralized Aggregation Algorithm in Federated Systems}}

\author{\textbf{Sumit Chongder}}
\affil{\small Department of Computer Science and Engineering, School of Computing, \\ Maharashtra Institute of Technology - Art, Design and Technology University, Pune, India \\ \textit{Email: sumitchongder960@gmail.com}}

\date{}

\begin{document}

\maketitle

\begin{abstract}
In recent years, the landscape of federated learning has witnessed significant advancements, particularly in decentralized methodologies. This research paper presents a comprehensive comparison of Centralized Hierarchical Federated Learning (HFL) with Decentralized Aggregated Federated Learning (AFL) and Decentralized Continual Federated Learning (CFL) architectures. While HFL, in its centralized approach, faces challenges such as communication bottlenecks and privacy concerns due to centralized data aggregation, AFL and CFL provide promising alternatives by distributing computation and aggregation processes across devices. Through evaluation of Fashion MNIST and MNIST datasets, this study demonstrates the advantages of decentralized methodologies, showcasing how AFL and CFL outperform HFL in precision, recall, F1 score, and balanced accuracy. The analysis highlights the importance of decentralized aggregation mechanisms in AFL and CFL, which effectively enables collaborative model training across distributed devices. This comparative study contributes valuable insights into the evolving landscape of federated learning, guiding researchers and practitioners towards decentralized methodologies for enhanced performance in collaborative model training scenarios.

\vspace{1em}
\end{abstract}

\section{Introduction}
Federated Learning (FL) represents a revolutionary approach to machine learning, transforming traditional centralized methodologies by enabling collaborative model training across multiple distributed devices \cite{Banabilah2022}. Initially conceptualized to address privacy concerns inherent in centralized data aggregation, FL has evolved into various forms, including Centralized Hierarchical Federated Learning (HFL), Decentralized Aggregated Federated Learning (AFL), and Decentralized Continual Federated Learning (CFL) \cite{Qu2023}. These adaptations accommodate diverse applications and scalability requirements, leveraging the copious amount of data produced at the edge while prioritizing data privacy and security \cite{Wu2020}.

The evolution of FL has prompted advancements in aggregation algorithms, distinguishing it from traditional centralized approaches \cite{Kaur2023}. Unlike centralized aggregation, where data is gathered and analyzed in one place, FL employs distributed optimization algorithms to aggregate model updates while preserving data privacy at the local level \cite{Feng2023}. Recent trends in DFL emphasize scalable and efficient aggregation algorithms, such as gossip-based protocols and secure multi-party computation, ensuring robustness and scalability in large-scale distributed environments \cite{Beutel2020}. This paper aims to explore the fundamental principles, methodologies, and applications of FL, shedding light on its significance and potential implications for various industries. By examining the nuances of HFL, AFL, and CFL, we aim to offer an in-depth comprehension of FL's role in shaping the future of machine learning.

\subsection{Problem Statement and Contributions}
The research provides a comprehensive comparison among three prominent paradigms within federated learning: Hierarchical Centralized Federated Learning (HFL), Aggregate Decentralized Federated Learning (AFL), and Continual Decentralized Federated Learning (CFL). In this investigation, emphasis is placed on evaluating their efficacy and efficiency within the context of client-server architectures \cite{Moon2023}, \cite{Zhang2023}. The study leverages two widely recognized datasets, MNIST \cite{Lecun1998} and Fashion-MNIST \cite{Xiao2017}, as benchmarks for performance assessment.

A distinctive aspect of this study lies in its detailed comparative analysis of Hierarchical Federated Learning (HFL) and Decentralized Federated Learning (DFL) paradigms using real-world datasets. A thorough evaluation, based on various performance metrics, is presented, offering an understanding of the advantages and disadvantages of the federated learning methodologies \cite{Karimireddy2023}. By delivering a detailed knowledge of the effectiveness and productivity of federated learning paradigms, this research aims to make a substantial impact on the current discussions in the domain of Federated Learning.

\subsection{Terms and Terminologies}
The following terms and terminologies are discussed in reference to the performance of our work:

\subsubsection{Accuracy}
In Centralized FL and Decentralized FL strategies, accuracy represents the ratio of correctly identified cases to the overall number of Fashion MNIST and MNIST images that were categorized. The formula for accuracy is depicted in (1). This pivotal metric highlights the effectiveness of collaborative model training methodologies.
\begin{equation}
    \text{Accuracy} = \frac{\text{True Positives} + \text{True Negatives}}{\text{Total Samples}}
\end{equation}

\subsubsection{Precision}
In Centralized FL and Decentralized FL strategies, precision evaluates the quality of positive predictions among all positive predictions made for Fashion MNIST and MNIST images classified. The formula for precision is provided in (2).
\begin{equation}
    \text{Precision} = \frac{\text{True Positives}}{\text{True Positives} + \text{False Positives}}
\end{equation}

\subsubsection{Recall}
In Centralized FL and Decentralized FL strategies, recall measures the model's capacity to accurately detect true positives from the total number of actual positives in the classification of Fashion MNIST and MNIST images. The formula for recall is given in (3).
\begin{equation}
    \text{Recall} = \frac{\text{True Positives}}{\text{True Positives} + \text{False Negatives}}
\end{equation}

\subsubsection{F1 Score}
In Centralized and Decentralized FL strategies, the F1 score calculates the weighted average of precision and recall for the classified images in the Fashion MNIST and MNIST datasets and is calculated using (4). It balances both metrics and provides a single value to evaluate a model's overall performance.
\begin{equation}
    \text{F1 Score} = \frac{2 * \text{Precision} * \text{Recall}}{\text{Precision} + \text{Recall}}
\end{equation}

\subsubsection{Aggregation (using FedAvg Algorithm)}
In Centralized FL, aggregation refers to the process of combining local model updates from participating clients (devices) into a global model held by a central server. In Decentralized FL, aggregation is decentralized, where clients directly communicate and collaboratively aggregate their model updates without relying on a central server.
\begin{equation}
    \theta_{g} = \sum_{c=1}^{N} \frac{n_{c}}{N}\, \theta_{c}
\end{equation}
where $\theta_{g}$ is the global model parameter, $\theta_{c}$ is the local client model parameter, $n_{c}$ is the specific client, and $N$ is the total number of clients.

\subsubsection{Build Time}
In Centralized FL, the build time denotes the duration required to train a global model on the central server using aggregated updates from individual clients. In Decentralized FL, the build time is distributed across clients, as each client trains its local model independently. The overall build time depends on the parallelization and coordination among clients.
\[
\text{Build Time} = \text{End Time (model training)} - \text{Start Time (model training)}
\]

\subsubsection{Classification Time}
In Centralized FL, the classification time refers to the inference (prediction) time when using the global model managed by the central server. In Decentralized FL, the classification time refers to the inference time when using local models on individual clients.
\[
\text{Classification Time} = \text{End Time (model evaluation)} - \text{Start Time (model evaluation)}
\]

\subsection{Layout of Paper}
The paper is organized in the following manner: Section 1.4 presents a summary of the current research that is relevant to HFL, AFL, and CFL. Section 2 details the Materials and Methods employed in the study. Section 3 discusses the Experimental Results and offers an in-depth examination. The paper is concluded in Section 4, which also suggests directions for Future Research.

\subsection{Related Work}
Recent years have witnessed considerable progress in the domain of decentralized federated learning (DFL). The present study adds to a body of literature reviews by offering an exhaustive summary of the latest advancements, informed by an array of scholarly articles.

\paragraph{A. Challenges of Federated Learning}
Federated Learning (FL) is a highly relevant field of research but is challenged by communication issues and heterogeneities. One of the main hurdles is ensuring efficient communication across the peers in the federated network \cite{Li2020}. The distribution of data across multiple devices often results in non-independent and identically distributed (non-i.i.d) and unbalanced datasets \cite{Nguyen2022}. Managing heterogeneous systems within the same network is another significant challenge. Privacy concerns also come into play, necessitating the development of privacy-preserving methods \cite{Li2019}.

\paragraph{B. Applications and Advantages of Decentralized Federated Learning}
In their enlightening study, E. T. Martínez Beltrán et al. \cite{Beltran2022} explain the basic principles of decentralized FL, including its frameworks, evolving trends, and inherent challenges. Their work effectively demonstrates how decentralized FL operates across a network of devices, preserving data privacy while enabling collaborative model training.

\paragraph{C. Survey and Exploration of Diverse Scenarios in Decentralized Federated Learning}
In a comprehensive survey by Yuan et al. \cite{Yuan2023}, invaluable insights into the landscape and trajectories of DFL are provided. Their analysis covers communication overhead, model aggregation techniques, and privacy preservation mechanisms. Concurrently, research by J. S. Nair et al. \cite{Nair2022} delves into the practical implications of DFL in multi-robot scenarios.

\paragraph{D. Research Directions in Decentralized FL}
T. V. Nguyen et al.'s innovative approach \cite{Nguyen2022} to DFL, tailored for training on medical data that is widely dispersed, of inferior quality, and confidential, exemplifies pioneering solutions to pressing challenges.

\section{Research Method}
In this section, we delve into the architecture and algorithms underpinning both centralized and decentralized federated learning methodologies.

\subsection{Hierarchical Federated Learning (HFL)}
Figures 1 and 2 explain the functioning of hierarchical federated learning (HFL), a collaborative approach to distributed model training while ensuring the protection of data privacy and security. It begins by initializing the global model parameters, followed by iterative training rounds \cite{HaghighiFard2024}. At the global server, gradients are aggregated with updates from other groups \cite{Tian2022}. The aggregated parameters are then disseminated back to the groups for further refinement \cite{Long2023}.

\begin{figure}[H]
  \centering
  % Replace with your actual filename (png, jpg, or pdf)
  \includegraphics[width=0.85\textwidth]{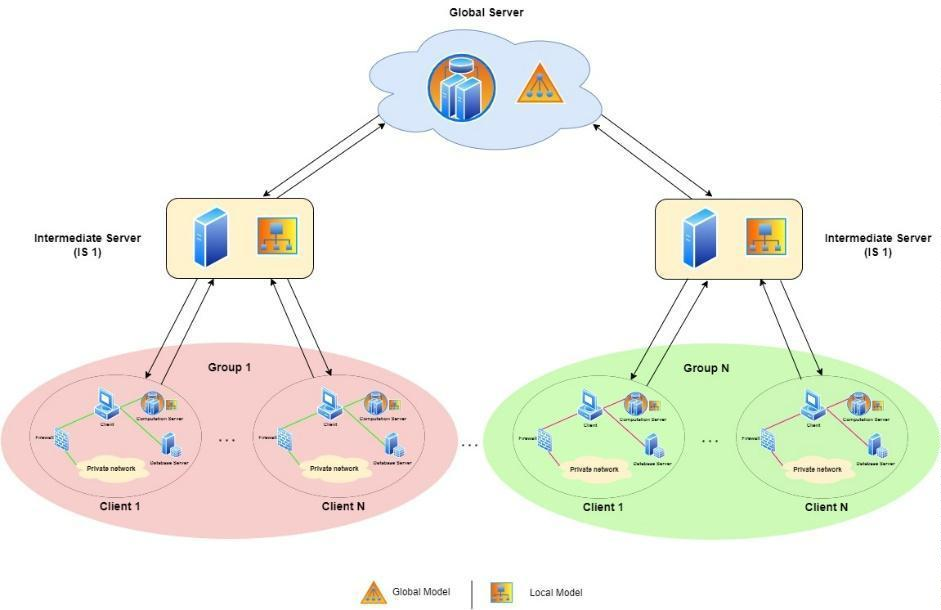}
  \caption{Architecture overview of hierarchical federated learning.}
  \label{fig:hfl_arch}
\end{figure}

\begin{figure}[H]
  \centering
  \includegraphics[width=0.4\textwidth]{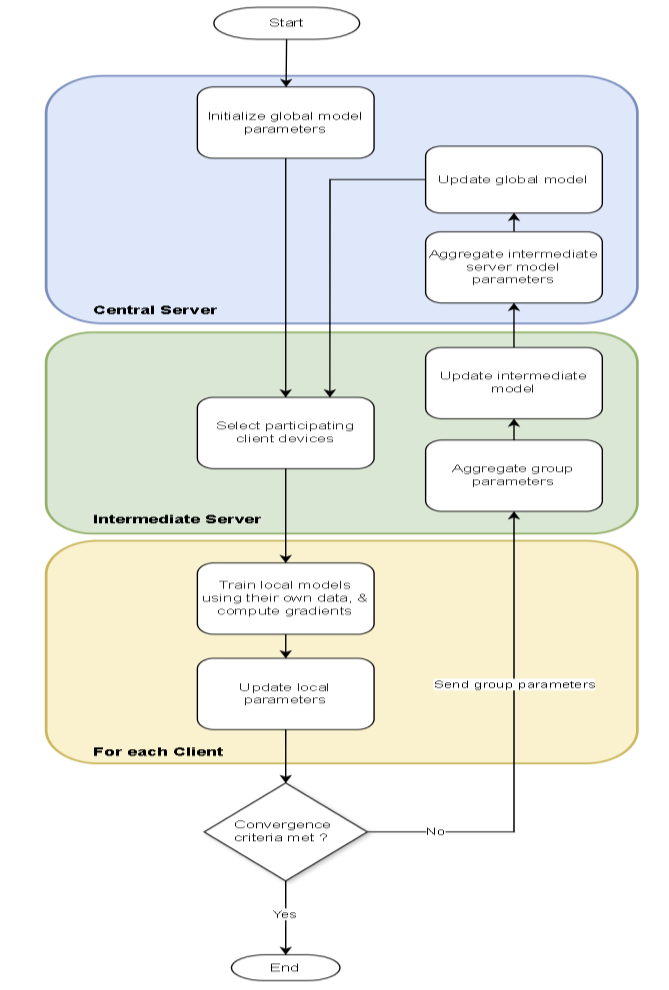}
  \caption{Flowchart of hierarchical federated learning.}
  \label{fig:hfl_flow}
\end{figure}

\subsection{Decentralized Aggregated Federated Learning (AFL)}
Figures 3 and 4 depict the operational framework of Decentralized Aggregate Federated Learning (AFL), which is a cooperative approach to machine learning in which various participants work together to train a global model over multiple rounds \cite{Wen2022}. AFL operates by selecting a subset of clients in each round, which then individually train the model locally for several epochs. Following local training, the models are aggregated using a designated aggregation function \cite{Malladi2021}.

\begin{figure}[H]
  \centering
  \includegraphics[width=0.7\textwidth]{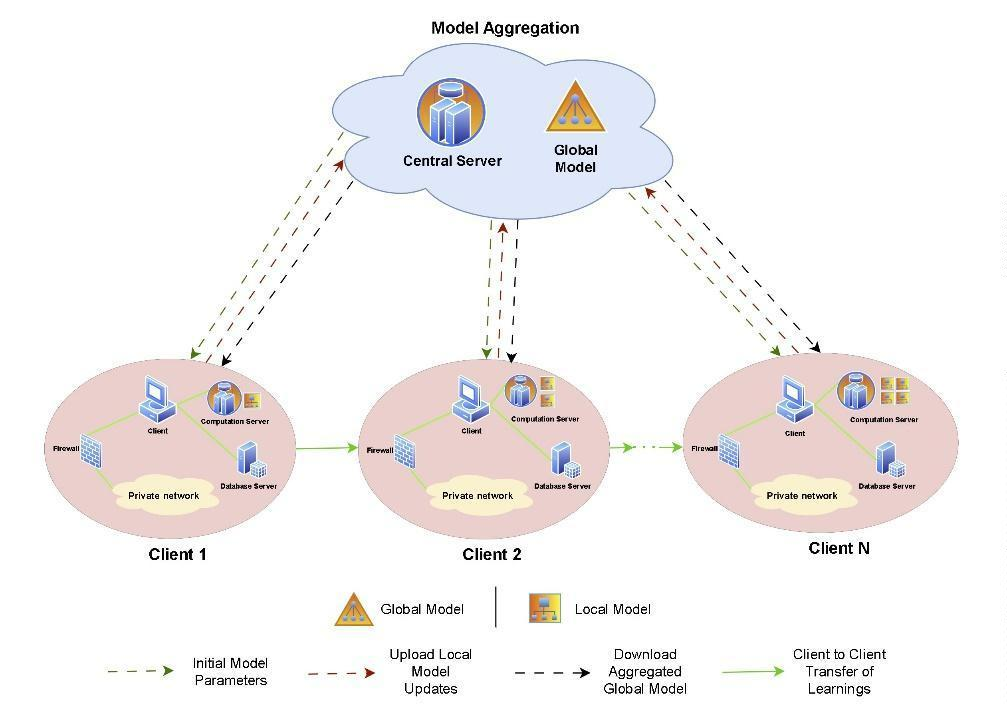}
  \caption{Architecture overview of decentralized aggregated federated learning.}
  \label{fig:afl_arch}
\end{figure}

\begin{figure}[H]
  \centering
  \includegraphics[width=0.6\textwidth]{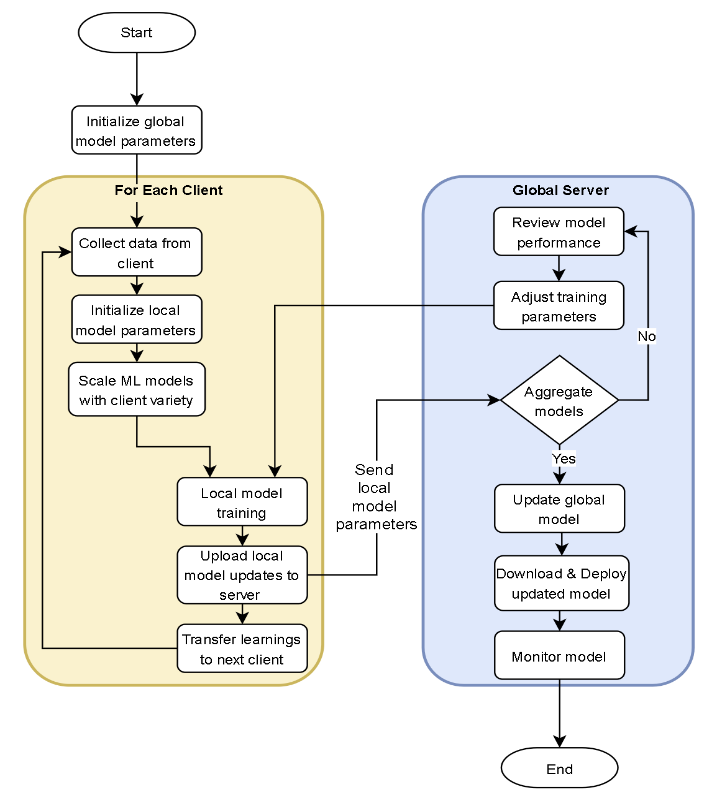}
  \caption{Flowchart of aggregated federated learning.}
  \label{fig:afl_flow}
\end{figure}

\subsection{Decentralized Continual Federated Learning (CFL)}
Figures 5 and 6 illustrate the operational workflow of Decentralized Continual Federated Learning (CFL), offering insights into its iterative training mechanism and collaborative model refinement across multiple client nodes. CFL is an iterative machine learning paradigm wherein numerous clients collaboratively refine a global model across multiple rounds \cite{Fu2023}. Crucially, CFL entails the continuous updating of local models, with these updates seamlessly integrated to adapt the parameters of the global model \cite{Xu2023}.

\begin{figure}[H]
  \centering
  \includegraphics[width=0.7\textwidth]{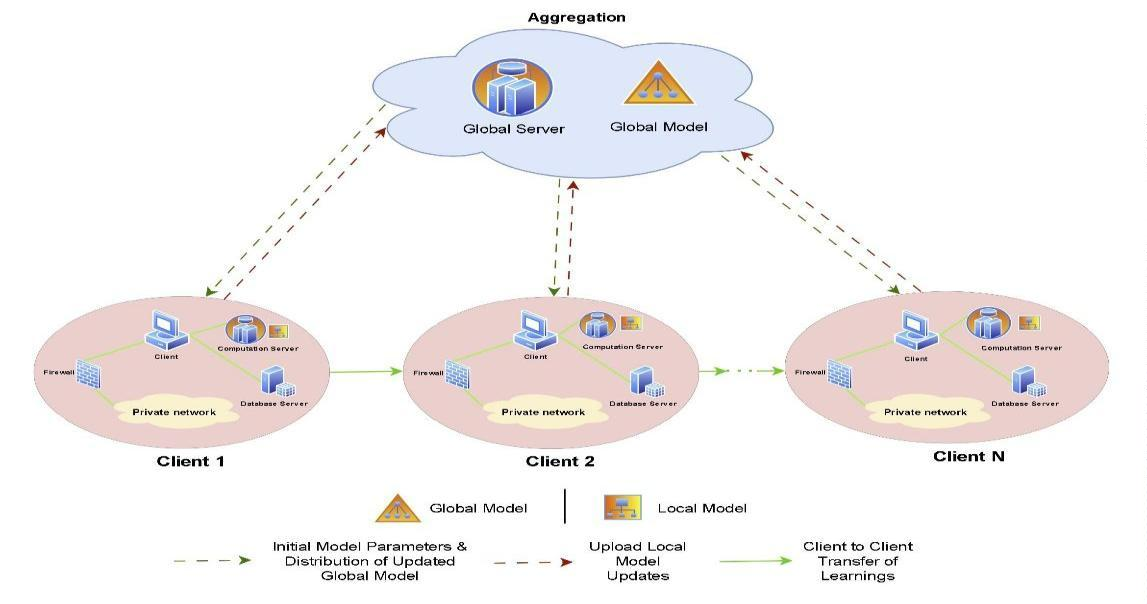}
  \caption{Architecture overview of decentralized continual federated learning.}
  \label{fig:cfl_arch}
\end{figure}

\begin{figure}[H]
  \centering
  \includegraphics[width=0.8\textwidth]{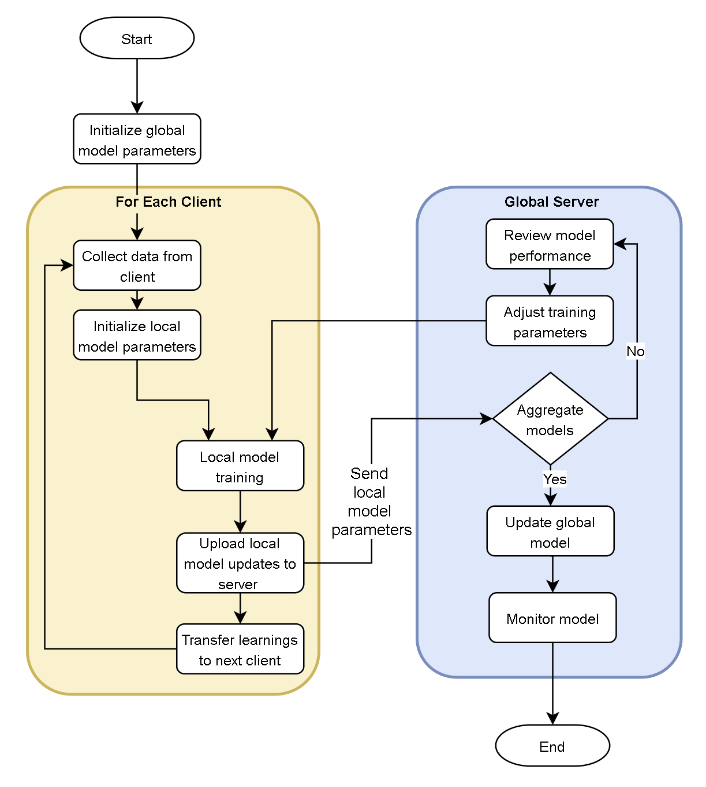}
  \caption{Flowchart of continual federated learning.}
  \label{fig:cfl_flow}
\end{figure}

\subsection{CNN Architecture}
In the experiment, an identical Convolutional Neural Network (CNN) is utilized due to its established efficacy in handling image data. The chosen CNN architecture includes three convolutional layers with 16, 12, and 10 filters, each with a filter size of 3x3. Two max-pooling layers are incorporated to downsample the feature maps. The ReLU activation function is employed in hidden layers.

\begin{figure}[H]
  \centering
  \includegraphics[width=0.8\textwidth]{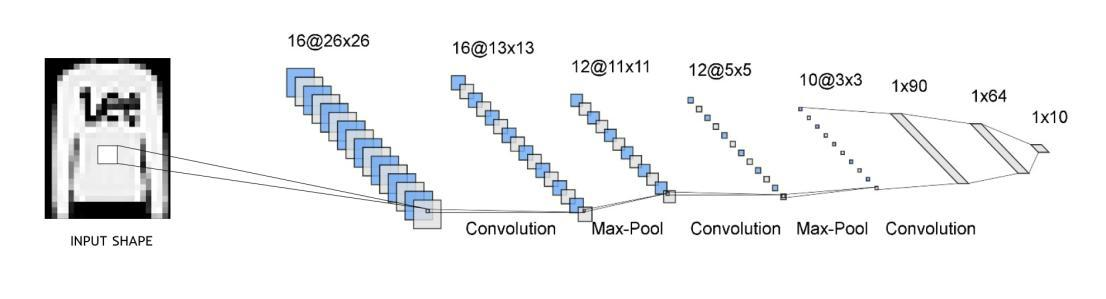}
  \caption{Proposed CNN architecture layers.}
  \label{fig:cnn_arch}
\end{figure}

\section{Results and Discussion}
The experimental outcome was achieved using tools such as Python 3.11.4, TensorFlow 2.16.1, Anaconda 2023.03, Jupyter Notebook v7.0.6, and Keras 3.1.1.

\subsection{Introduction to the Dataset}
Experiments are conducted on two publicly accessible datasets: MNIST \cite{Lecun1998} and Fashion MNIST \cite{Xiao2017}. MNIST comprises handwritten digits (0-9) of size 28x28 pixels. Fashion MNIST consists of grayscale images depicting various clothing items from ten different classes.

\subsection{Design of Experiment}
A data distribution that follows the Independent and Identically Distributed (IID) principle is implemented. The distribution is illustrated in Figure 8.

\begin{figure}[H]
  \centering
  \includegraphics[width=0.8\textwidth]{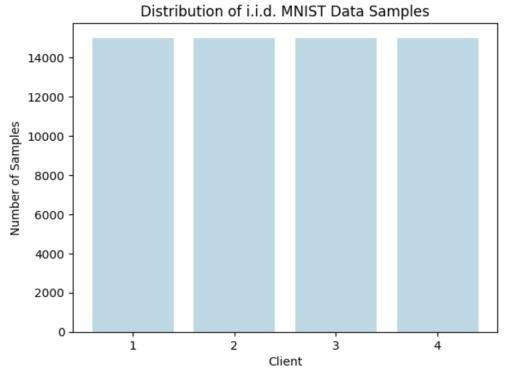}
  \caption{Sample of i.i.d. data from one of the clients.}
  \label{fig:iid_data}
\end{figure}

This study compared various Federated Learning (FL) paradigms. The models were trained using the training data, and their parameters were adjusted based on the evaluation inferred using the validation data. Table \ref{tab:perf_comp} presents the comparison of performance measures including Training accuracy, Testing accuracy, Build time, and Classification time.

\begin{table}[H]
\centering
\caption{Comparing the accuracy and time of the federated learning environments}
\label{tab:perf_comp}
\resizebox{\textwidth}{!}{%
\begin{tabular}{@{}lcccccccc@{}}
\toprule
 & \multicolumn{4}{c}{\textbf{Fashion MNIST}} & \multicolumn{4}{c}{\textbf{MNIST}} \\
\cmidrule(lr){2-5} \cmidrule(lr){6-9}
\textbf{Environment} & \textbf{\begin{tabular}[c]{@{}c@{}}Training\\ Acc.\end{tabular}} & \textbf{\begin{tabular}[c]{@{}c@{}}Testing\\ Acc.\end{tabular}} & \textbf{\begin{tabular}[c]{@{}c@{}}Build\\ Time (s)\end{tabular}} & \textbf{\begin{tabular}[c]{@{}c@{}}Class.\\ Time (s)\end{tabular}} & \textbf{\begin{tabular}[c]{@{}c@{}}Training\\ Acc.\end{tabular}} & \textbf{\begin{tabular}[c]{@{}c@{}}Testing\\ Acc.\end{tabular}} & \textbf{\begin{tabular}[c]{@{}c@{}}Build\\ Time (s)\end{tabular}} & \textbf{\begin{tabular}[c]{@{}c@{}}Class.\\ Time (s)\end{tabular}} \\ \midrule
Hierarchical FL & 0.85 & 0.41 & 86.11 & 0.57 & 0.93 & 0.60 & 88.26 & 0.55 \\
Aggregated FL & 0.93 & 0.70 & 55.33 & 0.47 & 0.95 & 0.72 & 54.02 & 0.52 \\
Continual FL & 0.95 & 0.88 & 80.07 & 0.31 & 0.96 & 0.98 & 79.02 & 0.29 \\ \bottomrule
\end{tabular}%
}
\end{table}

Based on the observed experimentation results, the decentralized FL approach exhibits superior performance compared to centralized FL methods across both Fashion MNIST and MNIST datasets. The accuracy and loss graph of the testing and training phase during the iterations is shown in Figure 9 (Fashion MNIST) and Figure 11 (MNIST).

\begin{figure}[H]
  \centering
  \includegraphics[width=0.9\textwidth]{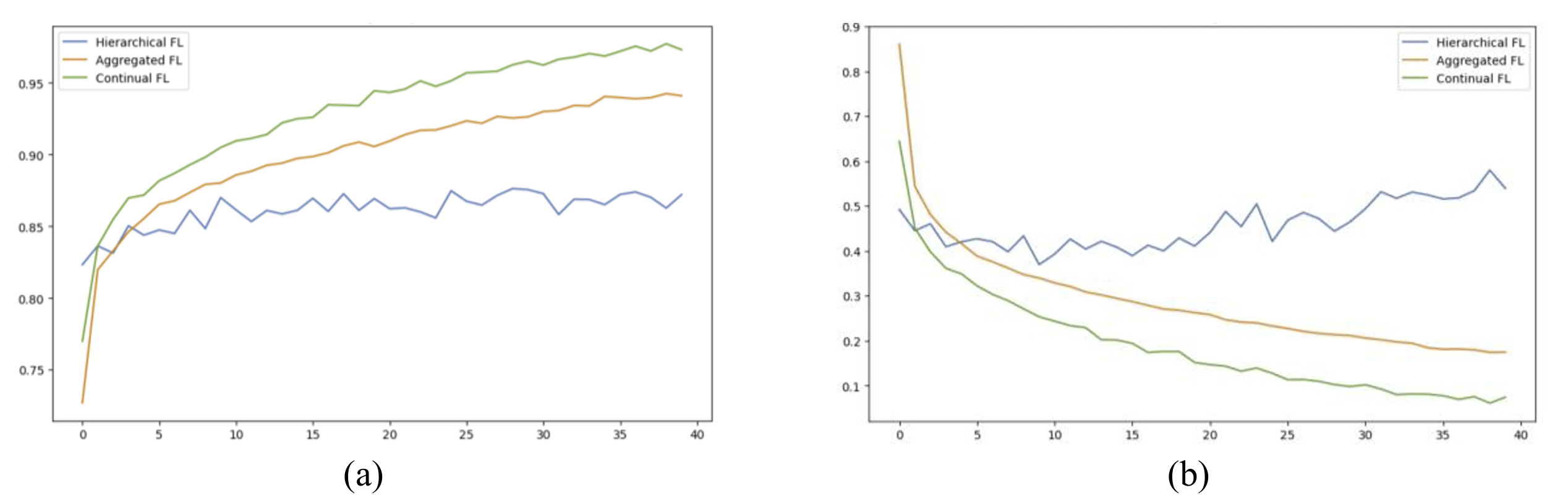}
  \caption{Training accuracy and loss comparison of the federated learning paradigms (Fashion MNIST): (a) Accuracy comparison and (b) Loss comparison.}
  \label{fig:fmnist_loss}
\end{figure}

To assess the effectiveness of the proposed approach, the confusion matrix is employed to depict the model's classification performance (Figure 10 for Fashion MNIST and Figure 12 for MNIST).

\begin{figure}[H]
  \centering
  \includegraphics[width=0.9\textwidth]{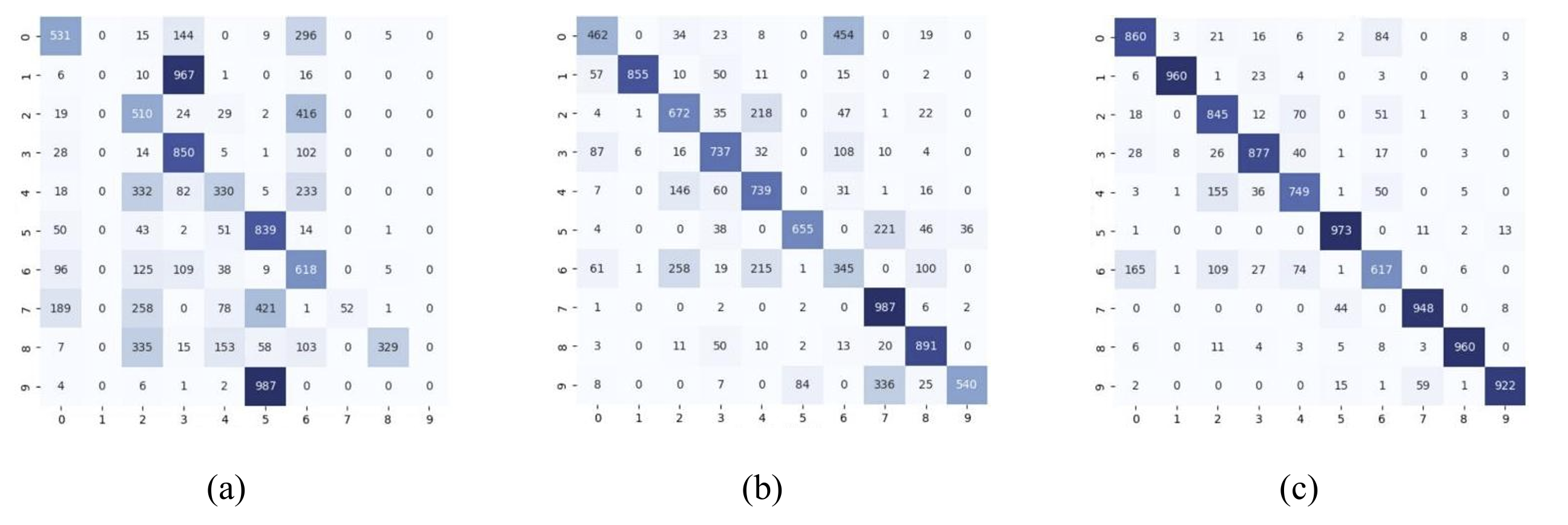}
  \caption{Confusion matrix of the federated learning paradigms (Fashion MNIST): (a) Hierarchical FL, (b) Aggregated FL, and (c) Continual FL.}
  \label{fig:fmnist_conf}
\end{figure}

\begin{figure}[H]
  \centering
  \includegraphics[width=0.9\textwidth]{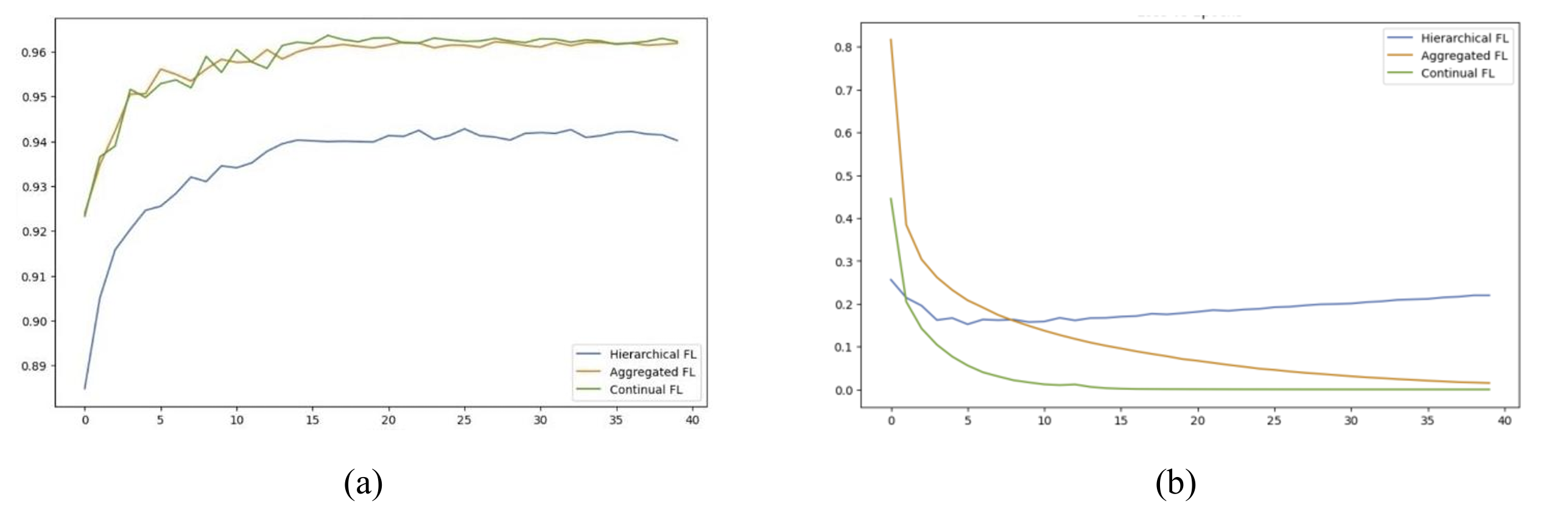}
  \caption{Training and loss comparison of the federated learning paradigms (MNIST): (a) Accuracy comparison and (b) Loss comparison.}
  \label{fig:mnist_loss}
\end{figure}

\begin{figure}[H]
  \centering
  \includegraphics[width=0.9\textwidth]{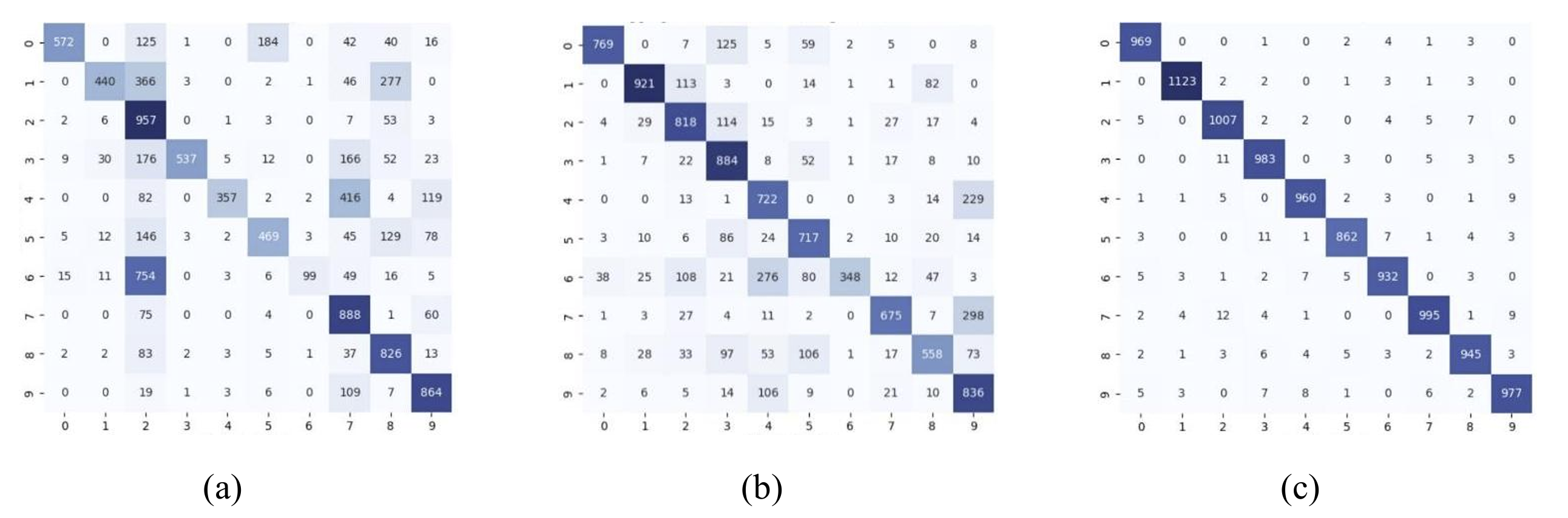}
  \caption{Confusion matrix of the federated learning paradigms (MNIST): (a) Hierarchical FL, (b) Aggregated FL, and (c) Continual FL.}
  \label{fig:mnist_conf}
\end{figure}

Table \ref{tab:res_comp} presents the comparison of performance measures (Precision, Recall, F1 Score) obtained from the proposed models.

\begin{table}[H]
\centering
\caption{Comparing the results of different federated learning environments}
\label{tab:res_comp}
\resizebox{\textwidth}{!}{%
\begin{tabular}{@{}lcccccccc@{}}
\toprule
 & \multicolumn{4}{c}{\textbf{Fashion MNIST}} & \multicolumn{4}{c}{\textbf{MNIST}} \\
\cmidrule(lr){2-5} \cmidrule(lr){6-9}
\textbf{Environment} & \textbf{Precision} & \textbf{Recall} & \textbf{F1 Score} & \textbf{Accuracy} & \textbf{Precision} & \textbf{Recall} & \textbf{F1 Score} & \textbf{Accuracy} \\ \midrule
Hierarchical FL & 0.41 & 0.33 & 0.40 & 0.41 & 0.75 & 0.60 & 0.59 & 0.60 \\
Aggregated FL & 0.71 & 0.68 & 0.68 & 0.69 & 0.76 & 0.72 & 0.72 & 0.72 \\
Continual FL & 0.88 & 0.87 & 0.86 & 0.88 & 0.98 & 0.98 & 0.98 & 0.98 \\ \bottomrule
\end{tabular}%
}
\end{table}

\subsection{Discussion and Analysis}
This section evaluates the performance of both centralized and decentralized FL strategies. Our findings demonstrate that decentralized FL approaches, such as Aggregated and Continual FL, consistently outperform the centralized approach in terms of accuracy (Figure 13).

\begin{figure}[H]
  \centering
  \includegraphics[width=0.9\textwidth]{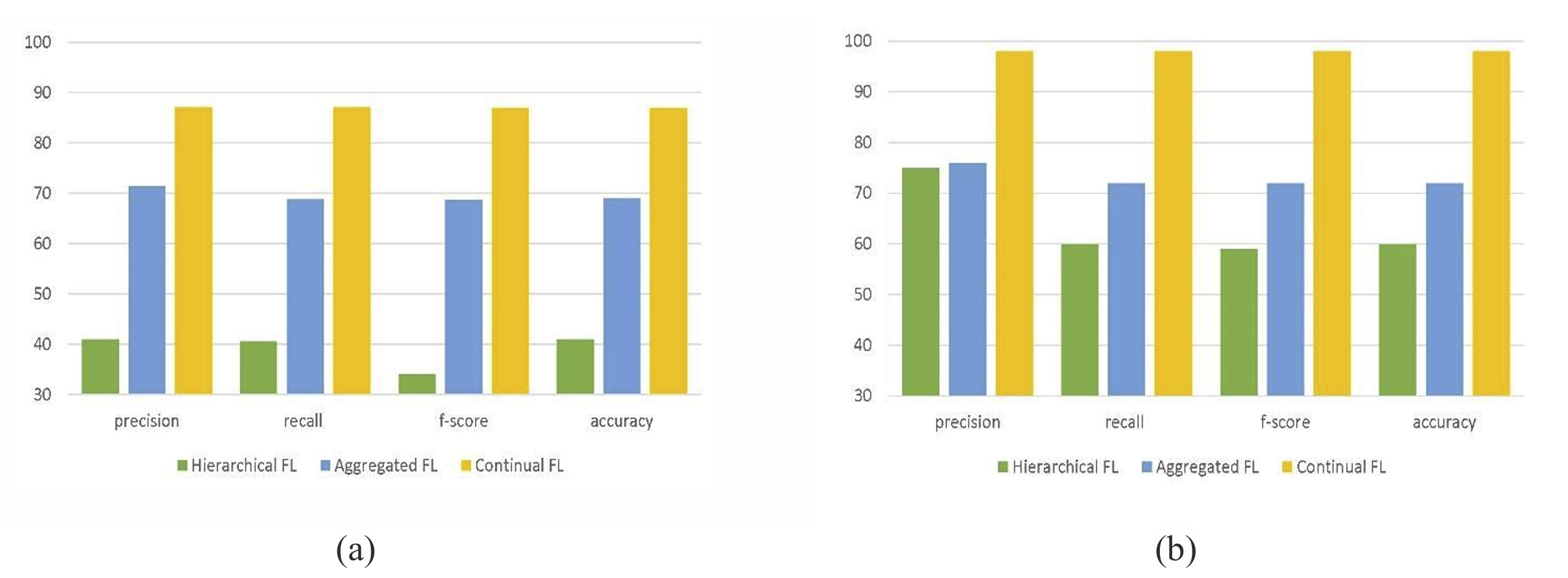}
  \caption{Accuracy of FL paradigms on two datasets: (a) Fashion MNIST and (b) MNIST.}
  \label{fig:acc_chart}
\end{figure}

In addition, the findings suggest that while Aggregated FL demonstrates the shortest build time across both datasets, it comes at the expense of training and testing accuracy compared to Continual FL. Furthermore, in terms of classification time, Continual FL showcases a notable advantage over both Hierarchical FL and Aggregated FL (Figure 14).

\begin{figure}[H]
  \centering
  \includegraphics[width=0.9\textwidth]{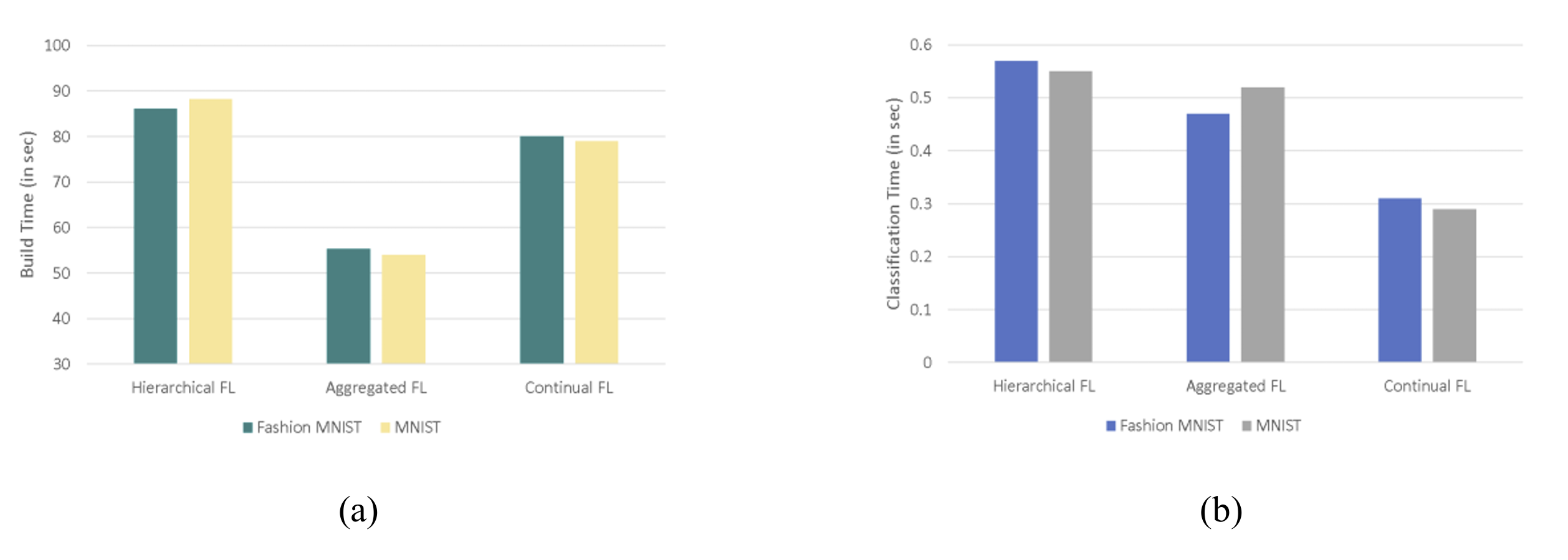}
  \caption{Computational efficiency of FL paradigms: (a) Build Time and (b) Classification Time.}
  \label{fig:eff_chart}
\end{figure}

\section{Conclusion \& Future Work}
This research examines various Federated Learning (FL) strategies in private data environments. It identifies the decentralized Continual FL method as exceptionally promising, surpassing centralized FL strategies in accuracy rates on validation and testing datasets, while also exhibiting reduced build and classification times.

Some potential directions for future work involve:
\begin{enumerate}
    \item \textbf{Exploring Data Distribution Combinations:} In DFL, a promising area is the systematic study of diverse data distributions.
    \item \textbf{Heterogeneity and Scalability:} Addressing the inherent heterogeneity within Decentralized Federated Learning (DFL) setups is crucial.
\end{enumerate}

\end{document}